\documentstyle[aaspp4,psfig,11pt]{article}

\def\unsetyr{\def\oyear{\relax}\def\cyear{\relax}\def\cyeara{a\relax}\def\cyearb{b\relax}\def\cyearc{c\relax}\def\cyeard{d\relax}}
\def\setyr{\def\oyear{(}\def\cyear{)}\def\cyeara{a)}\def\cyearb{b)}\def\cyearc{c)}\def\cyeard{d)}}
\unsetyr
\def\jcite#1{\setyr\cite{#1}\unsetyr}

\def\rmmat#1{{\hbox{\rm #1}}}
\def\rmscr#1{\rmmat{\scriptsize #1}}
\newcommand{\be}{\begin{equation}}
\newcommand{\ee}{\end{equation}}
\newcommand{\bt}{\begin{table} \begin{center}}
\newcommand{\et}{\end{center} \end{table}}
\newcommand{\ba}{\begin{eqnarray}}
\newcommand{\ea}{\end{eqnarray}}
\newcommand{\ie}{{\it i.e.~}}
\newcommand{\eg}{{\it e.g.~}}
\newcommand{\cf}{{\it c.f.~}}
\def\p{\partial}
\def\d{{\rm d}}

\def\dd#1#2{\frac{\d #1}{\d #2}}
\def\pp#1#2{\frac{\p #1}{\p #2}}
\newcommand{\comment}[1]{\relax}
\def\eqref#1{Equation~(\ref{eq:#1})}

\def\figref#1{Figure~\ref{fig:#1}}

\begin{document}
\newcommand{\bfi}{{\bf B}} \newcommand{\efi}{{\bf E}}
\newcommand{\lel}{{\lambda_e^{\!\!\!\!-}}}
\newcommand{\me}{m_e}
\newcommand{\mcs}{{m_e c^2}}
\def\ho{{\hat {\bf o}}}
\def\hm{{\hat {\bf m}}}
\def\hx{{\hat {\bf x}}}
\def\hy{{\hat {\bf y}}}
\def\hz{{\hat {\bf z}}}
\def\hom{{\hat{\mathbf{\omega}}}}
\def\hr{{\hat {\bf r}}}
\def\omv{\mathbf{\omega}}

\title{The High-Energy Polarization-Limiting Radius of Neutron Star
 Magnetospheres I --  Slowly Rotating Neutron Stars}
\author{Jeremy S. Heyl$^{\star}$\footnote{Chandra Postdoctoral Fellow},
Nir J. Shaviv$^{\dag}$\footnote{Current Address: Racah Institute of Physics,  
Hebrew University, Jerusalem 91904, Israel}
, and Don Lloyd$^{\star}$}
\affil{$^{\star}$
Harvard College Observatory, MS-51, 
60 Garden Street, Cambridge, Massachusetts 02138, United States}
\affil{$^{\dag}$
Canadian Institute for Theoretical Astrophysics, 
University of Toronto, 60 St.~George Street, Toronto, Ontario M5S 3H8,
Canada}

\begin{abstract}
  In the presence of strong magnetic fields, the vacuum becomes a
  birefringent medium. We show that this QED effect decouples the
  polarization modes of photons leaving the NS surface.  Both the
  total intensity and the intensity in each of the two modes is
  preserved along a ray's path through the neutron-star
  magnetosphere.  We analyze the consequences that this effect has on
  aligning the observed polarization vectors across the image of the
  stellar surface to generate large net polarizations. Counter to
  previous predictions, we show that the thermal radiation of NSs
  should be highly polarized even in the optical.  When detected, this
  polarization will be the first demonstration of vacuum
  birefringence. It could be used as a tool to prove the high magnetic
  field nature of AXPs and it could also be used to constrain physical
  NS parameters, such as $R/M$, to which the net polarization is
  sensitive.
\end{abstract}

\section{Introduction}

 The thermal radiation of isolated NS stars has the potential of
 teaching us much about the properties of NSs. Its advantage over
 non-thermal emission (in radio, optical, X-rays, and gamma-rays) is
 that the theory behind the emission is significantly better
 understood and the radiation actually comes from the surface of the
 compact object. Because the thermal emission is expected to be
 intrinsically polarized, more information could potentially be
 learned by the detection and analysis of polarization measurements.
 Recent observations with the {\em ROSAT}, {\em ASCA}, {\em Chandra}
 and {\em XMM-Newton} missions have shown that some of these sources
 are bright enough to be potential candidates for X-ray polarimetry in
 future missions. Moreover, the thermal radiation of some of the
 isolated NSs can even be detected in optical wavelengths. Thus, it is
 worthwhile understanding how this polarization is generated and
 conserved, and what additional information can actually be extracted
 from its measurement.

 In the presence of strong magnetic fields, the opacity of ionized
 matter to the transfer of photons becomes polarization dependent
 (\cite{1974ApJ...190..141L,1979PhRvD..19.1684V}). This is chiefly
 because it is easier to scatter electrons in the direction along the
 magnetic field than it is in a perpendicular direction. Thus, the
 opacity of light rays with their electric vector polarized perpendicular
 to the magnetic field would be significantly reduced.   A typical photon
 with this polarization is emitted or scattered last deeper in the atmosphere 
 than one in the other mode.  The deeper regions of the atmosphere
 are hotter, so more flux emerges in this polarization state.
 Nearly complete polarization can result for the thermal emission
 (\cite{Kann75,polaratmos1,polaratmos2}).

 An observer will see photons originating from the entire surface of the
 NS hemisphere facing her. Thus, the different polarizations should be
 added together appropriately. If nothing happens to the
 photons and their polarization as they propagate from the NS surface,
 then the polarizations observed at infinity can be added rather
 simply, as was done by \jcite{Kann75} in a simple model for the
 atmosphere. \jcite{Pavl00Thermal} used a more realistic atmosphere
 and calculated the net observed polarization while taking the effects
 that GR has on the magnetic field and on light ray bending. In both
 cases, net polarizations of order 5\% to 30\% are obtained because
 the polarizations of the radiation arriving from different regions of
 the surface tends to cancel each other.

 The above analyses, however, did not consider the effects of
 QED-induced vacuum birefringence. When QED is coupled to strong
 magnetic fields, several interesting consequences are obtained. For
 example, it was shown by \jcite{polaratmos1} and \jcite{Mesz79} that
 QED has to be taken into account when calculating the appropriate
 opacity, especially near the cyclotron resonance. This is the case
 even though a priori it appears that the plasma effects should
 dominate.  More relevant to us is the fact that QED turns the vacuum
 into a birefringent medium when strong magnetic fields are present
 (\cite{QEDbire1,QEDbire2}).

 In a series of recent papers, we have examined several consequences
 of vacuum birefringence in neutron-star magnetospheres.
 When the fields are significantly stronger than the critical QED field of
 $B_{QED}= 4.4 \times 10^{13}$~G then the index of refraction of one
 polarization state can be significantly different from unity and
 magnetic lensing can result (\cite{Shav98lens}). The main result of
 this lensing effect is that the effective surface area of the NS as
 measured by the two polarization states is different.   

 When weaker fields are present, the birefringence can still have
 interesting implications.  At a particular frequency, the vacuum will
 only decouple the polarization modes out to a particular distance
 from the surface of the star.   Up to this radius, radiation
 polarized perpendicular to the magnetic field will remain
 perpendicular to the local direction of the magnetic field even if
 the direction of the field changes along the path.
 If the modes only begin to mix at a
 significant fraction of the distance to the light cylinder, even if
 the intrinsic polarization at the surface is constant over energy,
 photons of different energies will exhibit different directions of
 polarization after passing through the magnetosphere
 (\cite{Heyl99polar}) and a circular component of the polarization
 will develop.  Similar effects arise at lower frequencies, when
 plasma birefringence is considered (\cite{Chen79,Barn86}). 

 In \jcite{Heyl01qed}, we showed that QED birefringence is also
 important for the polarization evolution close to the NS.  When it is
 properly taken into account, a very large net polarization should be
 observed. This is counter to previous predictions (\eg
 \cite{Pavl00Thermal}). As a result, larger polarization signals will
 be observable which will allow more information to be extracted from
 the observation of the thermal radiation.  Measurement of the high
 polarization will also serve as the first direct evidence of the
 birefringence of the magnetized vacuum due to QED and a direct probe
 of the behavior of the vacuum at magnetic fields of order of and
 above the critical QED field of $B_{QED} = 4.4 \times
 10^{13}$~G. This should be contrasted with the decades of Earth based
 experiments which have not succeeded thus far in detecting the vacuum
 birefringence induced by strong magnetic fields
 (\cite{Iaco79,Baka98,Rizz98,Nezr99}).

 We begin in \S\ref{sec:calculations} with the description of the
 physics needed to calculate the polarization to be observed at
 infinity. In \S\ref{sec:results}, we elaborate the results presented
 in \jcite{Heyl01qed} and build upon them to understand the
 observational signatures of vacuum polarization in rotating neutron
 stars.  \S\ref{sec:timeaverage} estimates the strength of the
 polarized signal averaged over the rotation of the star for the
 subset of radio pulsars for which we know the geometry of the dipole
 field.  We end in \S\ref{sec:discussion} with a discussion of the
 ramifications of this effect.

\section{Calculations}
\label{sec:calculations}

 Several ingredients are needed for the calculation of the net
 polarization to be observed at infinity. First, the structure of the
 magnetic field must be specified.  We assume that the magnetic field
 is a centered dipole.  Second, we need a model for the intrinsic
 polarization emitted by a magnetized atmosphere. For simplicity, we
 assume here that the atmospheres emit completely (linearly) polarized
 radiation. In a large frequency range, it is more than an adequate
 approximation because the effective temperature for the two
 polarizations will be markedly different if high magnetic fields are
 present.  For example, at photon energies $E_\gamma$ much below the
 electron rest energy and cyclotron energy $E_{cyc,e}$, but much above
 the ion cyclotron energy, the typical degree of linear polarization
 $p_L$ should be $1-p_L \sim {\cal O}(E_\gamma / E_\rmscr{cyc,e})^2$,
 unless the angle between the magnetic field ${\bf B}$ and the photon
 wavevector ${\bf k}$ is very small (\cite{Pavl00Thermal}).

 To calculate the observed polarization at infinity, we need to
 calculate the trajectories of the light rays, which due to GR are
 bent. Along these trajectories, we have to solve for the evolution of
 the polarization. This will be dominated by the vacuum birefringence.

\comment{
\subsection{Magnetic Field Structure}

 The structure of the magnetic field outside the star may be
 calculated from the magnetic scalar potential ($V$).  \jcite{Wald72}
 found that for a general sum of multipoles, the exterior field in the
 Schwarzschild spacetime is
\be
 V = \sum_{lm} q_{lm} f_l(r) Y_{lm} (\theta,\phi),
\ee
 where
\be
 f_l(r) = -\frac{(2l+1)!}{2^l (l+1)! \, l! \, M^{l+1}} \left (r - 2 M
 \right ) \dd{}{r} Q_l \left ( \frac{r}{M} -1 \right ).
\ee
 $Q_l(x)$ is the Legendre function of the second kind.

\paragraph{Dipole fields:}
 If we specialize to $l=1$, we obtain the result for a dipole field
\be
 {\vec B} = \frac{\mu}{r^3} \left [ \left ({\hat m} \cdot {\hat r}
 \right ) \left ( 2 F(u) + \psi(u) \right ) {\hat r} - \psi(u) {\hat
 m} \right ],
\label{eq:Bgr}
\ee
 where $u=2M/r$ and the auxiliary functions were calculated by
 \jcite{Ginz65}:
\ba
 F(u) &=& -\frac{3}{u^3} \left [ \ln ( 1 - u ) + u + \frac{1}{2} u^2
 \right ], \\ \psi(u) &=& \frac{3}{u^2} \left [ \frac{1}{1-u} +
 \frac{2}{u} \ln ( 1 - u ) + 1 \right ] \left ( 1 - u \right )^{1/2}.
\ea
 Far away from the star, $u\rightarrow 0$ and $F,\psi \rightarrow 1$,
 and we recover the standard result from magnetostatics and can
 identify $\mu$ with the magnetic moment measured by a distant
 observer.  The field at the magnetic pole of the star is given by
\be
 B_p = \frac{2 \mu}{R^3} F \left ( \frac{2 M}{R} \right );
\ee
the magnetic field is strengthened by the curvature of the
 spacetime near the star.
}

\subsection{Photon Trajectories}
\label{sec:photontraj}

In calculating the trajectories of the photons we assume that the
field is sufficiently weak such that the index of refraction for both
modes is approximately unity throughout the magnetosphere (\cf
\cite{Shav98lens}).  We also neglect the effect of the rotation of the
star on the spacetime surrounding it.

Without rotation, all planes that pass through the center of the star
are equivalent, so we can integrate the equations of motion for a
photon in the equatorial plane of the Schwarzschild metric.  The
trajectory is determined uniquely by the impact parameter $b$.
To integrate the polarization, we require the position of
the photon as a function of the proper length along its path.
\jcite{Misn73} give the differential equations for the trajectory
\ba
 \dd{t}{\lambda} &=& \left ( 1 - \frac{2M}{r} \right )^{-1} \\
 \dd{r}{\lambda} &=& \left [ 1 - \frac{b^2}{r^2} \left (1 -
 \frac{2M}{r} \right) \right ]^{1/2} \label{eq:drdlam} \\
 \dd{\theta}{\lambda} &=& 0 \\ \dd{\phi}{\lambda} &=& \frac{b}{r^2},
 \label{eq:dphidlam}
\ea
and also
\be
 \dd{l}{\lambda} = \left ( 1 - \frac{2M}{r} \right )^{-1/2}
\ee
where $M$ is the gravitational mass of the star.
Combining \eqref{drdlam} and \eqref{dphidlam} yields a separable
equation for $\phi(r)$ which is useful for quickly determining to
which part of the star a region of the image corresponds
(\cite{Page95})
\be
 \phi - \phi_0 = x \int_0^{M/R} \left [ \left ( 1- \frac{2 M}{R}
 \right ) \left ( \frac{M}{R} \right )^2 - \left ( 1 - 2 u \right )
 u^2 x^2 \right ]^{-1/2} \d u,
\label{eq:GRbending}
\ee
where $x\equiv b/R_\infty$ and $R_\infty \equiv R \left (1-2M/R
\right)^{-1/2}$ and $R$ is the circumferential radius of the star.

\subsection{Polarization Trajectories}

 The polarization lies in the plane perpendicular to the trajectory of
 the photon.  Unlike in flat spacetime, because the photon travels
 along a curved path, the orientation of this plane with respect to a
 distant observer necessarily varies along the path.  
 Under the assumption that space surrounding the
 neutron star is devoid of material and nongravitational fields, the
 polarization is constant, if it is defined in a basis consisting a
 vector in the plane of the trajectory and one perpendicular to that
 plane (\cite{1977MNRAS.179..691P}).

\begin{figure}
\plotone{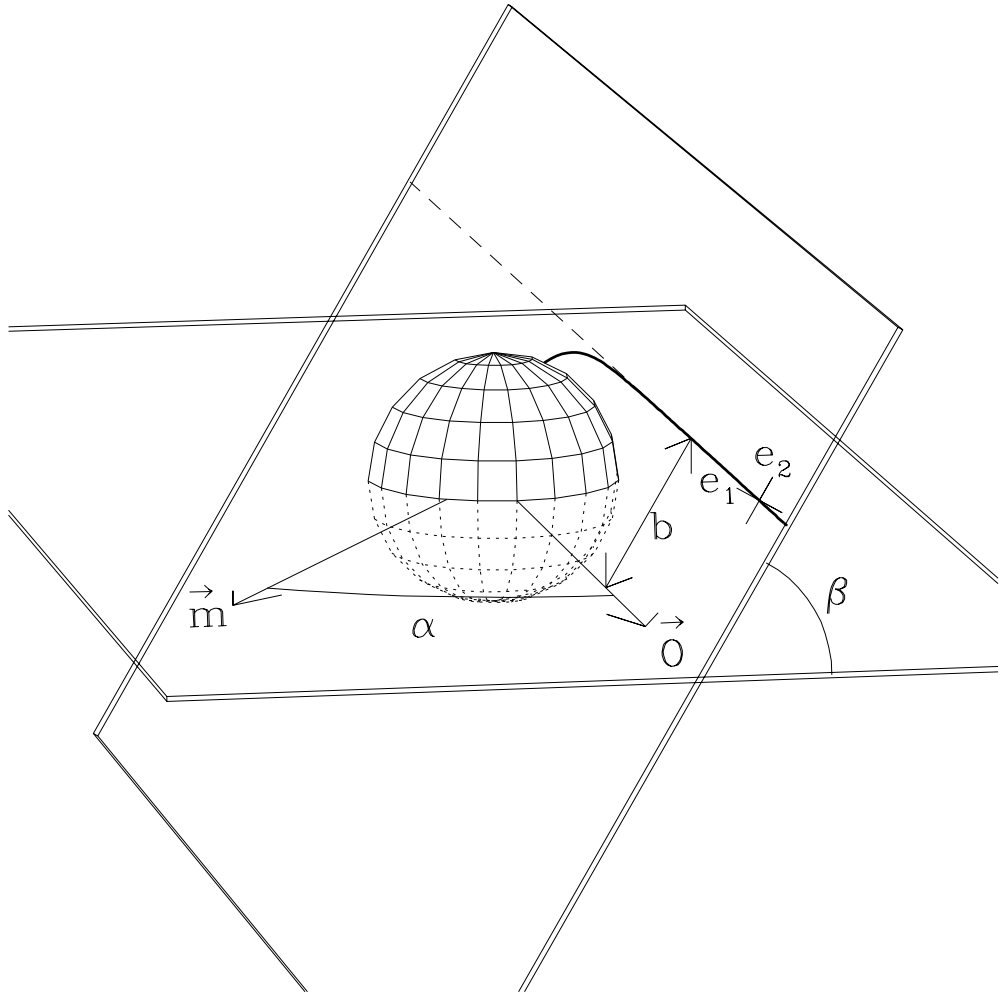}
\caption{The geometry of the photon trajectory and the polarization basis.}
\label{fig:geometry}
\end{figure}

 To naturally include the standard general relativistic result, we
 calculate the evolution of the polarization due to the birefringence
 of the vacuum in the aforementioned basis.  \figref{geometry} shows an example
 trajectory along with the polarization basis at a particular point.
 $\alpha$ is the angle between the magnetic dipole ($\vec m$) and the
 line of sight ($\vec O$), and $\beta$ is the angle between the
 trajectory plane and the ${\vec m}-{\vec O}-$plane.

 \jcite{Kubo83} find that the evolution of the polarization of a wave
 traveling through a birefringent and dichroic medium in the limit of
 geometric optics is given by
\newcommand{\hatom}{{\bf {\hat \Omega}}}
\be
\pp{\bf s}{l} = \hatom \times {\bf s} + 
\left ( {\bf {\hat T}} \times {\bf s} \right ) \times {\bf s},
\label{eq:sevol}
\ee
 where $l$ is the proper distance along the trajectory, ${\bf s}$ is
 the normalized Stokes vector (\cite{Jack75}), and $\hatom$ and ${\bf
 {\hat T}}$ are the birefringent and dichroic vectors.  The Stokes
 vector consists of the four Stokes parameters, $S_0, S_1, S_2$ and
 $S_3$.  The vector ${\bf s}$ consists of $S_1/S_0, S_2/S_0$ and
 $S_3/S_0$. The result was found by \jcite{Kubo83} for any dielectric
 medium and it was extended for a medium that is both dielectric and
 permeable by \jcite{Heyl99polar}.

 As \jcite{Heyl99polar} argue, QED decouples the polarization states
 in the vacuum for sufficiently strong fields.  Here we will restrict
 ourselves to fields substantially less than $B_\rmscr{QED} \approx
 4.4\times 10^{13}$~G.  A field of $10^{12}$~G is sufficient to
 decouple the polarization states at the surface of a neutron star for
 $\nu \gtrsim 10^{12}$~Hz.  A plasma with the Goldreich-Julian density
 decouples the polarization states of photons with $\nu \lesssim
 10^{14}$~Hz.  Here we will focus on ultraviolet through X-ray
 radiation, so the plasma contribution to the index of refraction may
 be neglected.  We will consider photons with $\nu \lesssim
 10^{14}$~Hz (\ie for which the plasma would be important) but we will
 also neglect the plasma contribution so we can connect our results
 with those of \jcite{Pavl00Thermal} who neglect the birefringence of
 the magnetosphere entirely.

 If one neglects the plasma and takes the weak-field limit, the
 dichroic vector vanishes and the magnitude of the birefringent vector is
\be
 \left |\hatom \right | = \frac{2}{15} \frac{\alpha_\rmscr{QED}}{4
 \pi} \frac{\omega}{c} \left ( \frac{B_\perp}{B_\rmscr{QED}}
 \right)^2,
\ee
 and it points in the direction of the projection of the magnetic
 field onto the Poincar\'e sphere.  $B_\perp$ is the strength of the
 magnetic field perpendicular to the direction of the photon's
 propagation.  We assume that the magnetic field is a centered dipole
 and we neglect the distortion of the magnetic field due to general
 relativity.  For the masses and radii that we are considering, the
 perturbation to the field strength is at most a factor of two and the
 change in the direction of the field is less than 5$^\circ$
 throughout (\cite{Ginz65}).  Both the value of the
 polarization-limiting radius and the emergent flux depend weakly on
 the strength of the magnetic field at the surface -- both increase as
 $B^{0.4}$; therefore, this simplification does not have an important
 effect on the results.

\jcite{Heyl99polar} find that the polarization states are decoupled as
long as the gradient of the index of refraction is not too
large, specifically
\be
 \left |\hatom \left ( \frac{1}{|\hatom|} \pp{ |\hatom |}{l} \right
 )^{-1} \right | \gtrsim 0.5~.
\label{eq:couplingratio}
\ee
 For radial trajectories this yields
\be
 r \lesssim r_\rmscr{pl} \equiv \left ( \frac{\alpha_\rmscr{QED}}{45}
 \frac{\nu}{c} \right )^{1/5} \left ( \frac{\mu}{B_\rmscr{QED}} \sin
 \alpha \right )^{2/5} \approx 1.2 \times 10^{7} \mu_{30}^{2/5}
 \nu_{17}^{1/5} \left ( \sin \alpha \right)^{2/5} \rmmat{cm}.
\label{eq:rpl}
\ee
 where $\mu$ is the magnetic dipole moment of the star and $\nu$ is
 the frequency of the photon.  Also, $\mu_{30} = \mu / (10^{30} \rmmat{ G cm}^3)$ and
 $\nu_{17}=\nu/10^{17} \rmmat{Hz}$.

\begin{figure}
\plottwo{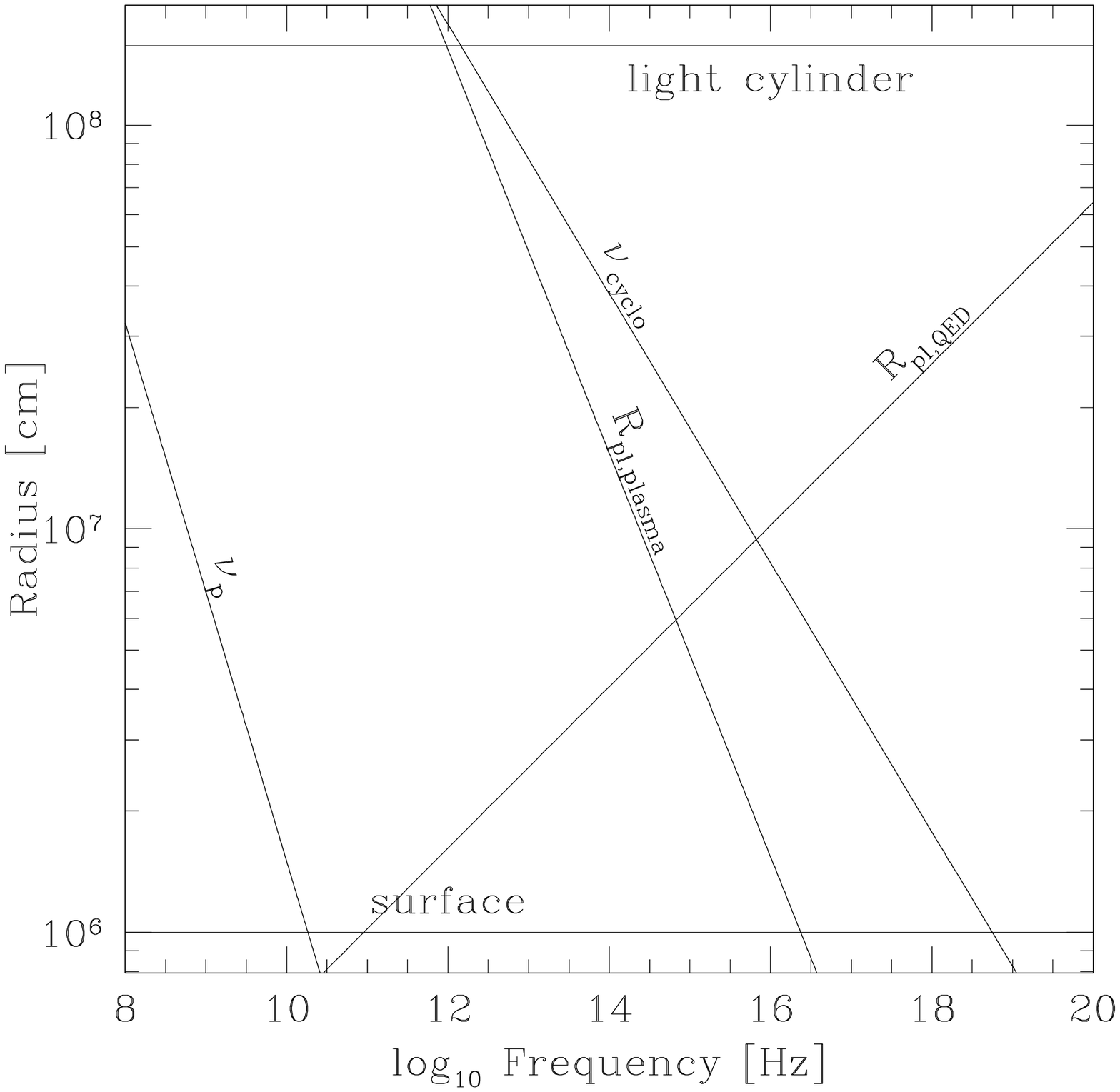}{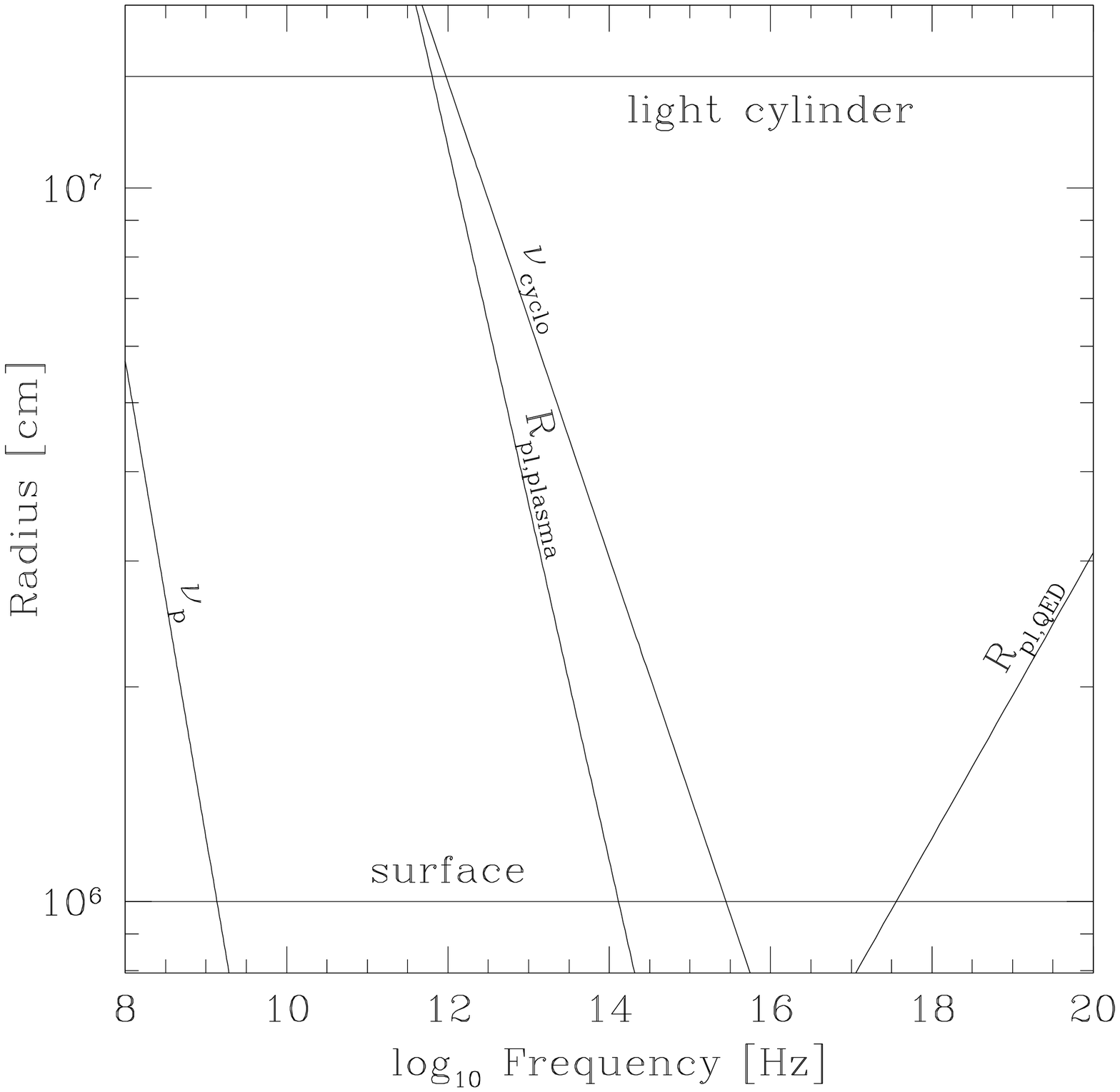}
 \caption{Polarization-limiting radius as a function of frequency.  The left
 panel is for a neutron star with $B=2\times 10^{12}$~G and $P=33$~ms.
 The right panel is for $B=10^9$~G and $P=3$~ms.}
\label{fig:couplerad}
\end{figure}

 If this condition is not met, the polarization remains constant, \ie
 the polarization modes are coupled.  \figref{couplerad} depicts the
 radii within which either the plasma or the vacuum effectively
 decouples the polarization modes in the magnetosphere -- we have
 assumed that the plasma density is given by the \jcite{Gold69}
 result.  At distances closer to the star than the
 polarization-limiting radius ($r_\rmscr{pl}$), the polarization of
 the radiation remains in one of the two polarization modes of the
 strongly magnetized plasma or vacuum.

 If the entire surface emits in one polarization mode, \ie the surface
 emission is initially fully polarized, the radiation will remain in
 that mode until the polarization-limiting radius, so one can estimate
 the observed extent of the polarization geometrically by calculating
 the solid angle subtended by the image of the surface of the star at
 the polarization-limiting radius.  As the angular size of the image
 at $r_\rmscr{pl}$ vanishes, the polarized fraction approaches unity.
 Numerical integration of the photon paths bears this out.

\section{Results}
\label{sec:results}

These results assume that $B \ll B_{QED}$ at the decoupling
radius. This is always the case for the photon energies of interest,
because even if the field on the surface is greater than the QED
value, as is the case in magnetars, the decoupling takes place far
enough from the surface such that the $r^{-3}$ term will make the
field significantly sub-critical.

\subsection{Effects along a trajectory}
\label{sec:trajectory}

\begin{figure}
\plotone{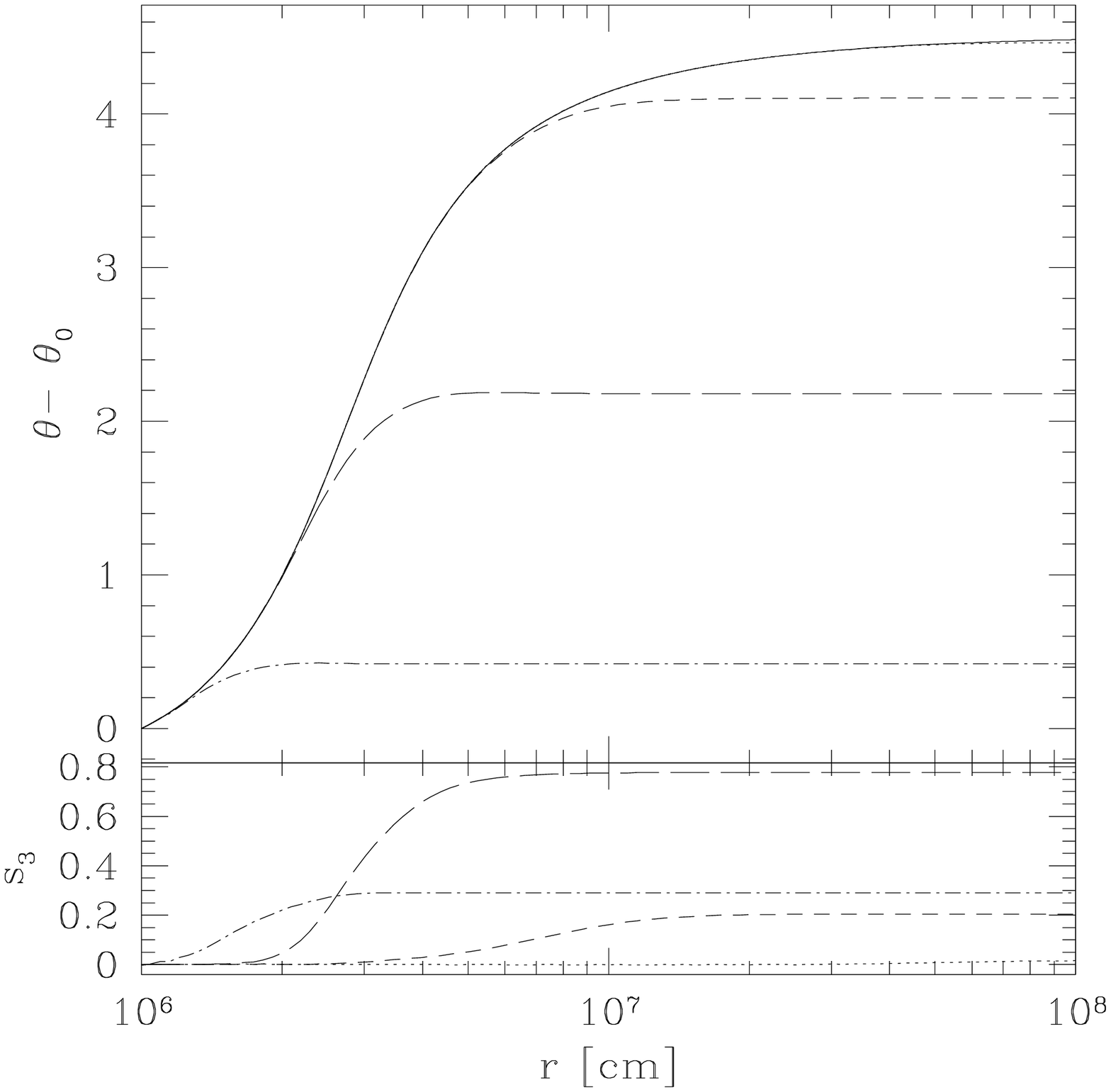}
 \caption{The evolution of the polarization for a particular ray as a
 function of frequency for $R=10^6$~cm, $M=2.1\times 10^5$~cm,
 $\alpha=30^\circ$, $\beta=148.5^\circ$ and $b=0.4 R_\infty$.  The
 solid curve traces the position angle of the birefringent vector
 $\hatom$, the dot-dashed line follows the polarization at $\nu
 \mu_{30}^2=10^{21}$~Hz, the long-dashed line at $10^{17}$~Hz, the
 short-dashed lines at $10^{15}$~Hz and the dotted line at
 $10^{13}$~Hz.}
\label{fig:polar_traj}
\end{figure}

 We begin by integrating the photon light trajectories for specific
 photons leaving the NS surface and following the evolution of their
 polarization. Because the QED vacuum is not dichroic by itself,
 eq.~(\ref{eq:sevol}) dictates that the amplitude of ${\bf s}$ does
 not change along a ray. In the course of evolution, however, the
 linear component of ${\bf s}$, \ie , the 1-2 components, may change
 direction, and the amount of circular component $s_3$ can change as
 well. The top part of figure \ref{fig:polar_traj} depicts the
 evolution of the angle of ${\bf s}$ in the 1-2 plane of the
 Poincar\'e Space, together with
 the angle of the birefringent vector $\hatom$ (determined by the magnetic
 field component perpendicular to the ray).

 As the particular photon leaves the surface, the magnetic field
 orientation rotates by about 2.2 radians which corresponds to 4.4
 radians in the 1-2 plane of the Poincar\'e Space. Because the coupling
 is weaker at lower frequencies, the lower frequency photons follow
 the direction of the magnetic field up to a smaller distance. Beyond
 the polarization-limiting distance, the polarization direction
 freezes. Its direction roughly corresponds to the direction of the
 magnetic field where the modes couple. Because modes couple
 gradually, the direction of the magnetic field can change during the
 coupling process. If the change in direction is rapid,
 a large circular polarization results. This is seen in the bottom
 part of figure \ref{fig:polar_traj}.  For the low frequency and high
 frequency photons, the coupling takes place before the magnetic field
 can significantly change or after it has stopped changing, so for
 these frequencies, the circular component obtained is small. For the
 intermediate frequency, which for $10^{12}$~G NSs corresponds to the
 optical or UV region, the coupling takes place while the direction of
 the magnetic field is changing and a large circular component is
 generated.

 At high frequencies (hard X-ray to $\gamma$-rays), the coupling can
 take place at a significant fraction of the light cylinder radius. In
 this case, it was shown by \jcite{Heyl99polar} that one should take
 into account the rotation of the NS.  Because the modes of higher
 frequencies couple farther from the NS, the directions of the
 rotating magnetic field at the polarization-limiting radii are
 different for different photon energies, such that phase leads
 between different wave bands can result. Because coupling takes place
 while the NS is rotating, circular polarization can again result.

\subsection{The Polarization ``Image'' of a NS}
\label{sec:image}

 The next step is to construct a polarization ``image'' of a NS. The
 apparent surface is projected onto a surface perpendicular to the
 NS-observer direction. The image is then divided into
 elements of equal solid angle.  Next, light rays are followed from
 each element of the apparent surface to the observer taking into
 account GR light bending (eq.~\ref{eq:GRbending}) and polarization
 evolution (eq.~\ref{eq:sevol}), as is described in
 \S\ref{sec:trajectory}.

\begin{figure}
\plottwo{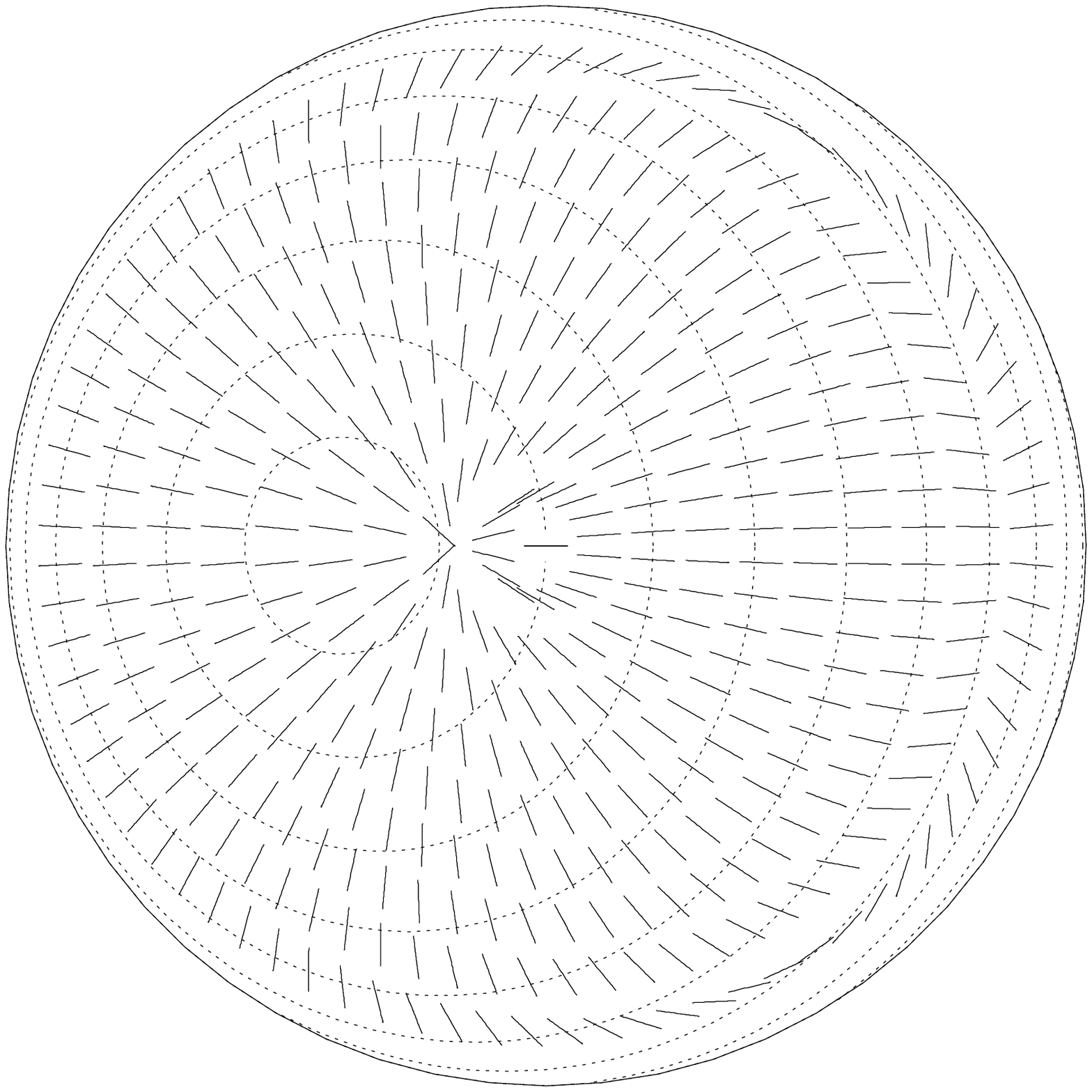}{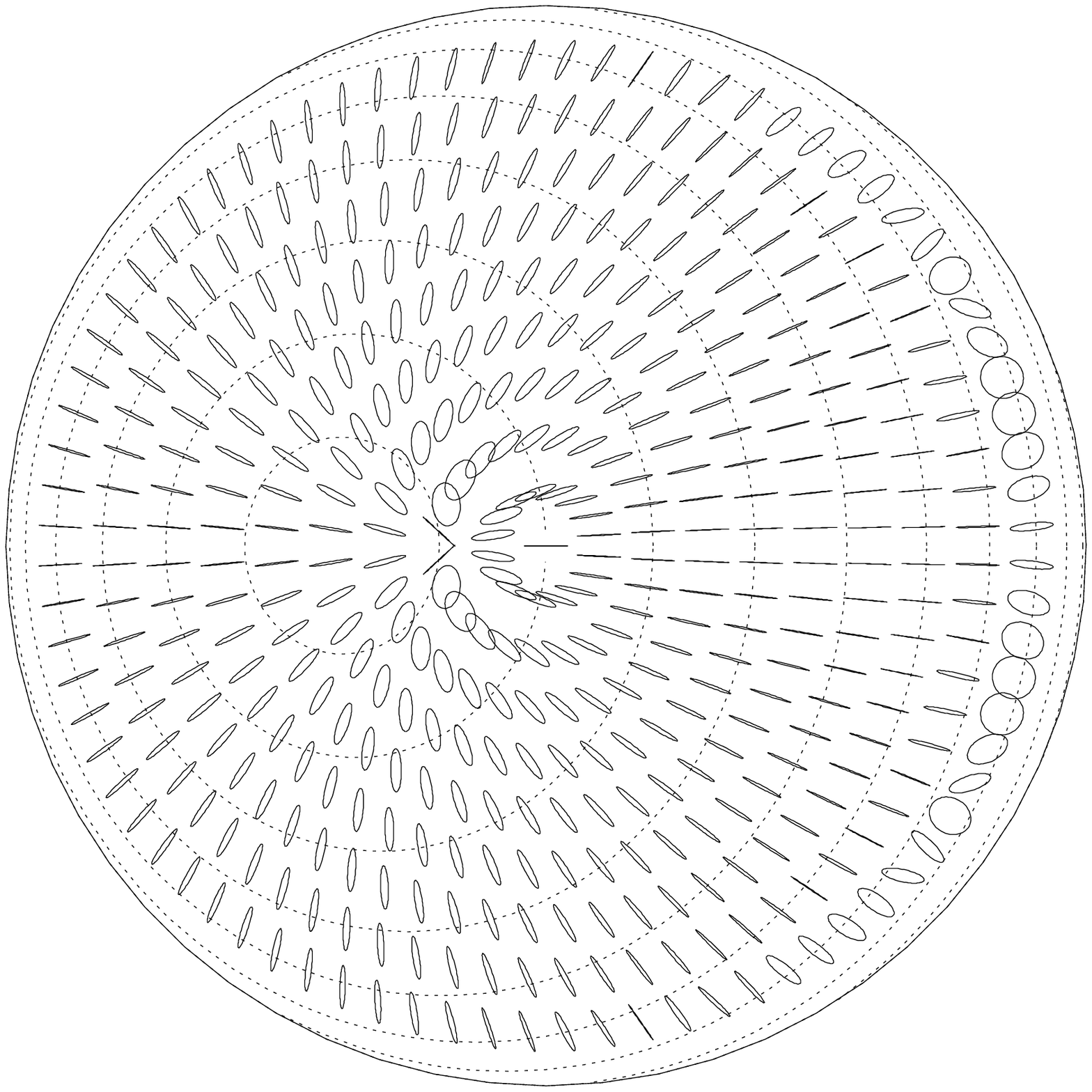} 

\plottwo{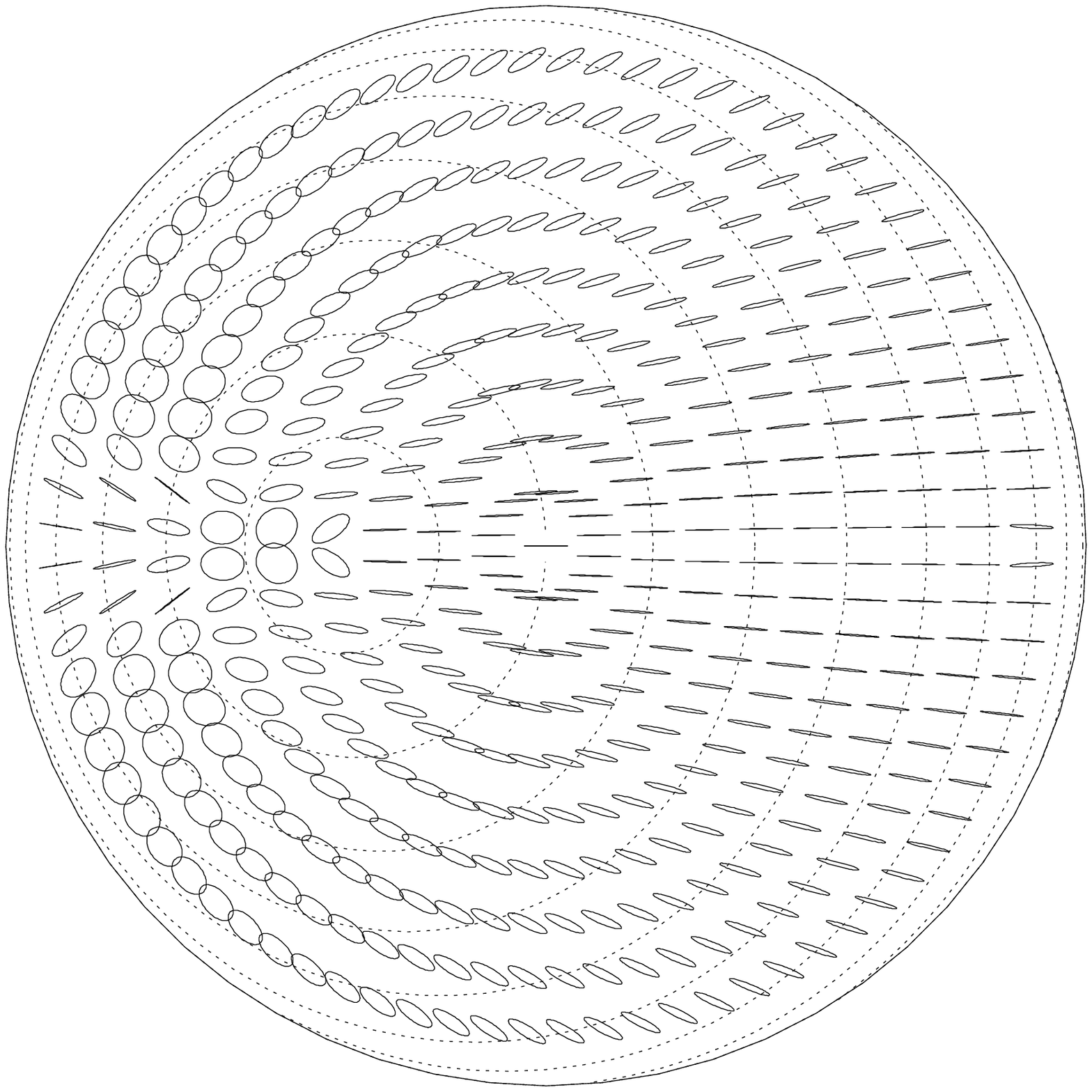}{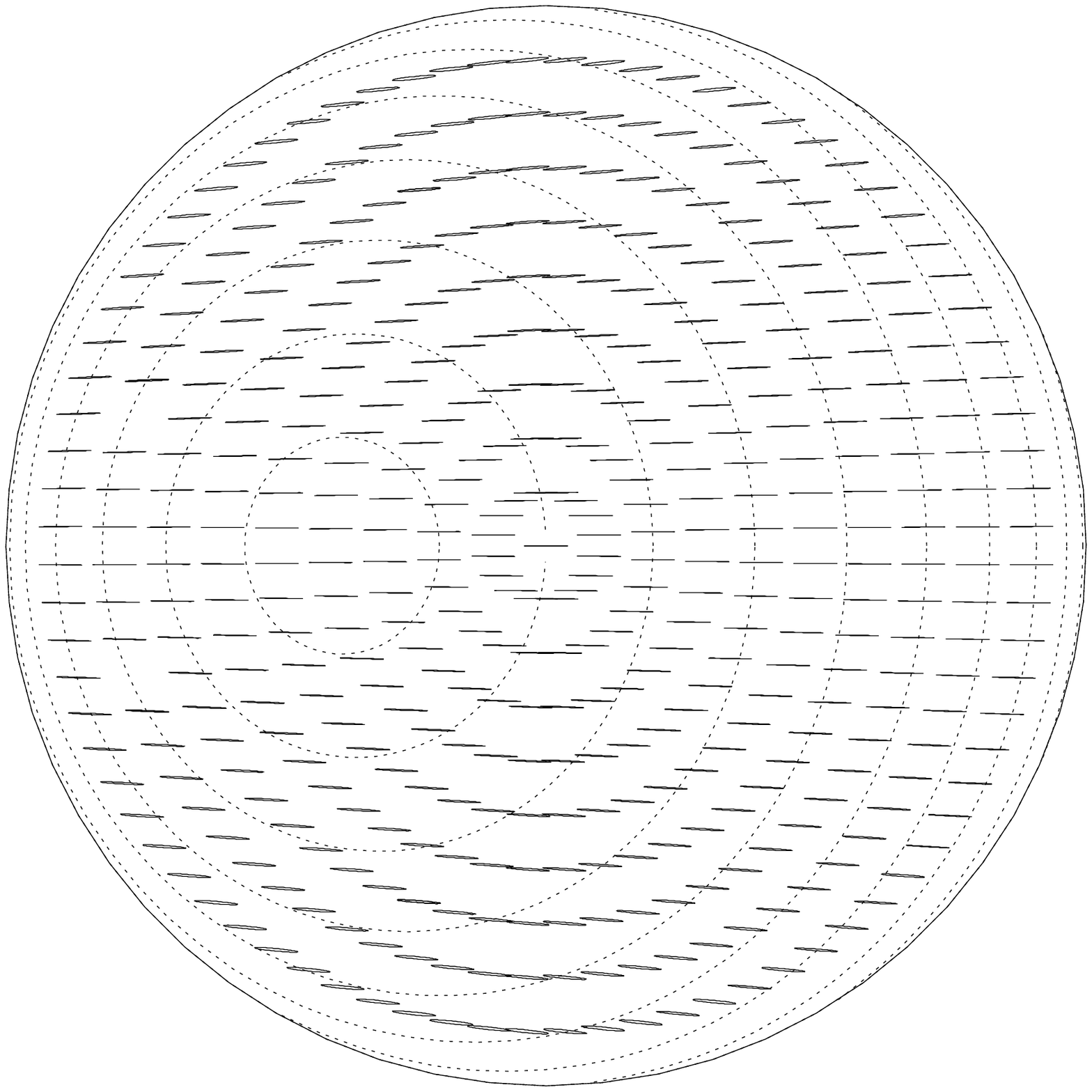}
 \caption{The observed polarization field for $R=10^6$~cm,
 $M=2.1\times 10^5$~cm and $\alpha=30^\circ$.  The upper-left and
 right panels are $\nu \mu_{30}^2=0$ and $10^{13}$~Hz. The lower-left
 and right panels are $\nu \mu_{30}^2=10^{17}$~Hz and $10^{21}$~Hz.}
\label{fig:pfield}
\end{figure}

 Typical results are portrayed in figure \ref{fig:pfield}, which shows
 the polarization observed at infinity overlaying the GR lensed image
 of the NS. The typical polarization is an ellipse. The major axis
 describes the direction of the linear polarization (in real space)
 while the ratio of the minor to major axis gives the amount of
 circular polarization. Not given in the figure is the sense of
 rotation of the circular polarization. From symmetry, one obtains
 that the circular components in the top half of the images are opposite to
 those in the bottom half. The same is true for the $s_2$ polarization
 which describes the $\pm 45^\circ$ polarization directions in real
 space. Unlike the $s_3$ antisymmetry however, the $s_2$ antisymmetry
 is apparent in the images.

 The anti-symmetry of the $s_2$ and $s_3$ components implies that when
 the polarizations from all the star will be added together, only a
 net $s_1$ component will result. Namely, the net polarization from
 the NS will be either in the direction of the magnetic dipole axis or
 perpendicular to it. This statement will not be true if the
 cylindrical symmetry is broken, either by the magnetic field, by
 rotation or by the atmospheric emission.

\subsection{The Net Polarization of a NS}

Once a polarization ``image'' is calculated, the net polarization seen
by an observer is found by integrating the intensity contributed by
each of the normal modes of the atmosphere to each of the observed
Stokes's parameters.  Here we will treat the simple case where the
intensity in one mode vanishes ($I_O=0$), and the intensity in the
other mode is isotropic ($I_X=$~constant).  In this case, the value of
$S_1/S_0$ is simply the mean value of $s_1$ evaluated over the
observed polarization field (\eg as depicted in \figref{pfield}).  For
this simple model, we denote $S_1/S_0$ summed over the entire image by
${\bar s}_1$.

\begin{figure}
\plottwo{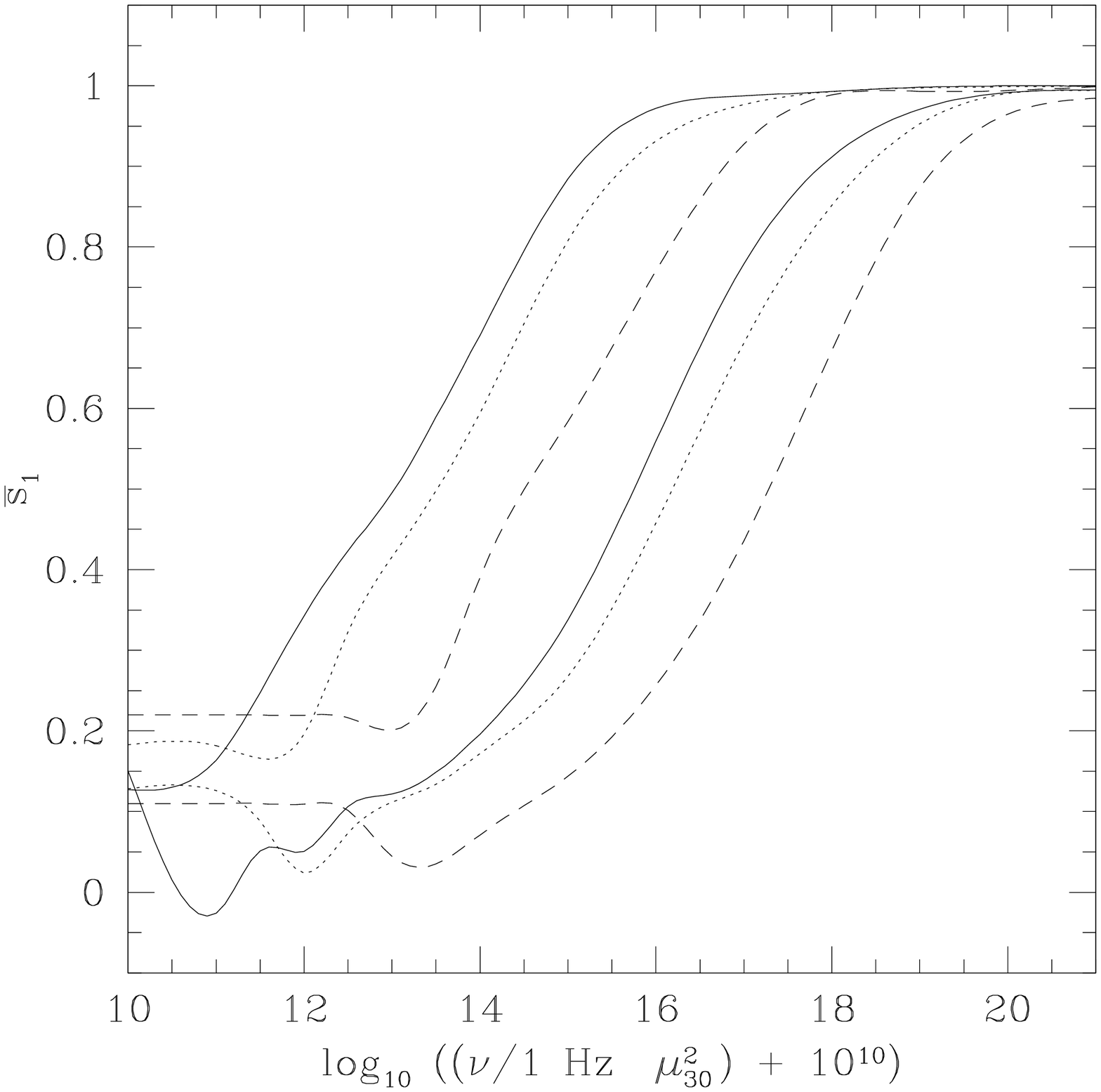}{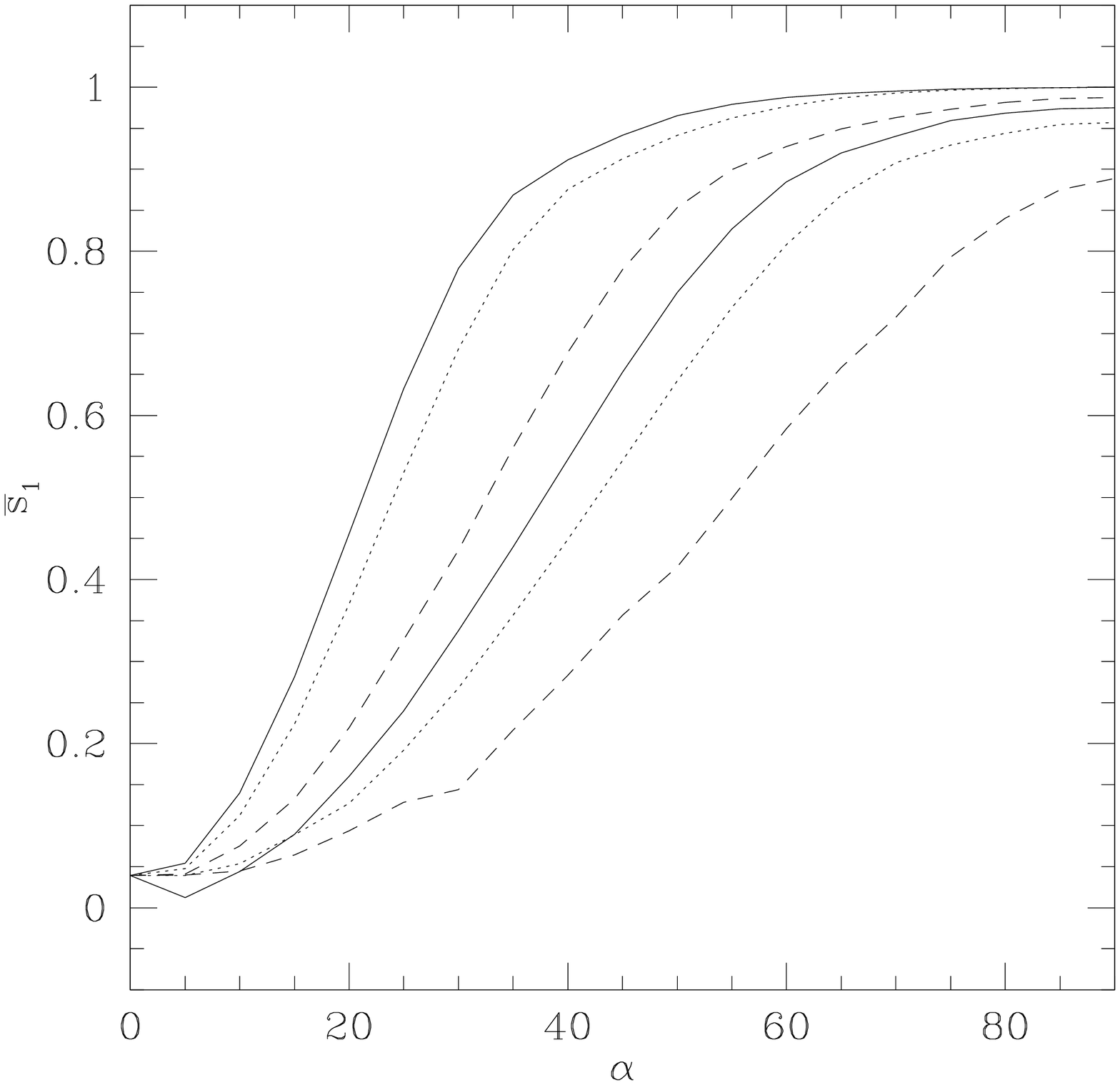} 
\caption{The left panel depicts the
polarized fraction as a function of $\nu\mu_{30}^2$ for $R=6, 10$ and
18~km in solid, dotted and dashed lines respectively and
$\alpha=30^\circ$ and $60^\circ$ for the lower and upper set of
curves respectively.  The right panel shows the polarized fraction as
a function of $\alpha$ for $R=6, 10$ and 18~km and $\nu
\mu_{30}^2=10^{15}$ (lower set) and $10^{17}$~Hz (upper set).  }
\label{fig:nus1}
\end{figure}

The results for ${\bar s}_1$ as a function of the magnetic field
strength and frequency $\nu\mu_{30}^2$ is given in the left panel of
figure \ref{fig:nus1}, for two inclination angles and three different
NS radii, 6~km, 10~km and 18~km. The right panel depicts the net
polarization ${\bar s}_1$ as a function of the angle between the line
of sight and the magnetic dipole moment for the three NS radii and two
frequencies.

\figref{nus1} depicts several important trends:
\begin{enumerate}
\item
Higher frequency radiation is more strongly polarized.
\item 
More strongly magnetized stars exhibit stronger polarization.
\item
As the line of sight approaches the direction of the dipole, the net 
polarization vanishes.
\item
At high frequencies, the emission from larger stars is {\em less} polarized.
The trend is reversed at low frequencies.
\end{enumerate}

 \eqref{rpl} predicts the first three of these trends directly.  As
 the frequency of the photon or the strength of the dipole moment
 increase, the polarization-limiting radius increases.  Also as $\sin
 \alpha$ increases, the polarization-limiting radius increases.  A
 larger value of $r_\rmscr{pl}$ results in a larger polarized fraction
 because the solid angle subtended by the bundle of rays that
 eventually reach the detector decreases with distance from the star.
 Over successively smaller solid angles, the magnetic field geometry
 appears successively more uniform, and the polarization from
 different regions of the star is added more coherently.

 The final trend requires a two-part explanation.  If $\nu \mu^2_{30}
 \sin^2 \alpha \gg 10^{12}$~Hz, then $r_\rmscr{pl} \gg R$, so the net
 polarization depends almost entirely on the angular size of the ray
 bundle.  Far from the surface of the star, the linear radius of the
 bundle is $b_\rmscr{max}$.  For a given mass, $b_\rmscr{max}$
 decreases with the radius of the star until it reaches a constant value
 for $R<3 M$; consequently, smaller stars have smaller bundles and
 larger net polarizations.

 For $\nu \mu^2_{30} \sin^2 \alpha \lesssim 10^{12}$~Hz, the
 polarization-limiting radius is comparable or smaller than the radius
 of the star.  In this regime, magnetospheric birefringence has little
 effect on the polarized image; therefore, in this regime, the results
 of \jcite{Pavl00Thermal} are obtained.  We see a larger fraction of
 the surface of more compact stars so the net polarization will
 decrease as $M/R$ increases, because the polarization is then added
 mostly incoherently.  For $R < 3.5 M$ we see the entire surface and
 for $R \leq 3 M$ we see an infinite number of images of the surface
 (\eg \cite{Page95}).

The paragraphs that follow examine the ramifications of these trends
in more detail. In particular, we calculate of the polarized light
curve of a neutron star and its average. We will continue in a
subsequent publication with predictions of the net polarization for a
realistic model of the emission from the surface of a neutron star.

\subsubsection{Polarization light curve of a NS}

 When an observer measures the polarization of a rotating NS, the
 amount and angle of polarization will generally vary because the
 rotation axis and magnetic axis are usually not aligned together.  We
 define the angular separation between the magnetic and rotational
 axes as $\gamma$. The magnetic inclination angle $i$ ($\equiv \pi/2 -
 \alpha$) can be related to the inclination above the rotational
 equator $i_r$ and the rotational phase $\phi$ between the last time
 the two axes coincided in the observer's meridional plane. The
 relation is
\be
 \sin i = \sin i_r \cos \gamma + \cos i_r \sin \gamma \cos \phi.
\ee
If we work in a coordinate system aligned with the rotational
$z$-axis, and a $y$-axis that is perpendicular to the plane
containing the line of sight and the $z$-axis, then the observer's
direction is:
\be
\ho = \cos i_r \hx + \sin i_r \hz.
\ee
If we use the rotational phase $\phi$ and the separation $\gamma$
between the two axes, the direction of the magnetic axis is:
\be
\hm = \sin \gamma \cos \phi \hx + \sin \gamma \sin \phi \hy + \cos
\gamma \hz.
\ee
 Using these relations, we can calculate the cosine and sine of twice
 the {\em apparent} angle $\psi$ that the magnetic axis makes with the
 $y$-axis. These are needed if we wish to known the direction of
 linear polarization. To do so, we project the polarization state
 $\left|S_1\right>$, in which the net polarization will be in (e.g.,
 \S\ref{sec:image}) onto the polarization states
 $\left|S_{O,1}\right>$ and $\left|S_{O,2}\right>$ of the
 observer. $\left|S_{O,1}\right>$ describes linear polarization in the
 observers $y$-axis and $\left|S_{O,2}\right>$ describes polarization
 in a direction rotated by $45^\circ$. The projections are
\ba
\left< S_{O,1} | S_1 \right> & = &\cos 2 \psi = 
{2 \left( (\hm - ( \hm \cdot \ho) \ho)\cdot \hy
 \right)^2 \over \left|\hm - ( \hm \cdot \ho) \ho) \right|^2} - 1 \\
 &=& { 2 \left( \sin \gamma \sin \phi \right)^2 \over 1 - \left( \cos
 \gamma \sin i_r + \cos i_r \sin \gamma \cos \phi \right)^2} - 1
 \equiv p_1 (\gamma,i_r,\phi),
\ea
and
\ba
 \left< S_{O,2} | S_1 \right> & = & \sin 2 \psi = {2 \left( (\hm - ( \hm
 \cdot \ho) \ho)\cdot \hy \right)\cdot \left( (\hm - ( \hm \cdot \ho)
 \ho)\cdot \hx \right) \over \left|\hm - ( \hm \cdot \ho) \ho)
 \right|^2} \\ &=& { 2 \sin \gamma \sin i_r \sin \phi \left( \cos \phi
 \sin \gamma \sin i_r - \cos \gamma \cos i_r \right) \over 1 - \left(
 \cos \gamma \sin i_r + \cos i_r \sin \gamma \cos \phi \right)^2}
 \equiv p_2 (\gamma,i_r,\phi).
\ea
 If the total net polarization observed at infinity $|\bar{s}|$ is
 calculated, the functions $p_{1,2} (\gamma,i_r,\phi)$ can now be
 used to find the actual polarization that will be measured if our
 polarizers are aligned with the $z$ axis or $45^{\circ}$ to it
 respectively.

\begin{figure}
\plotone{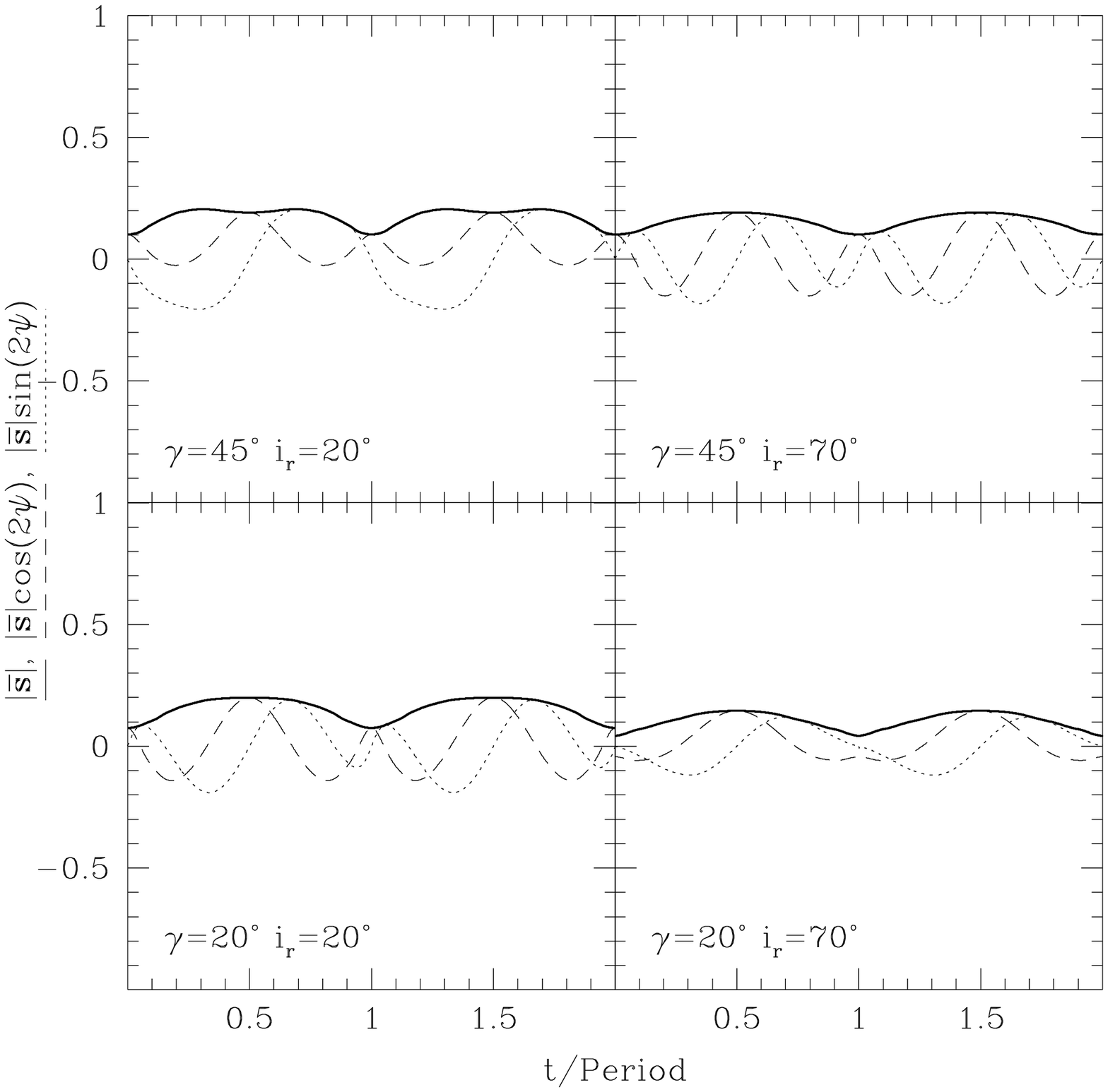}
 \caption{The polarized light curve for $\nu =0$ Hz and $R=10$ km
 (i.e., neglecting the effects that QED has on aligning the
 polarization). The solid line describes the total linear polarization.}
\label{fig:lc0}
\end{figure}

\begin{figure}
\plotone{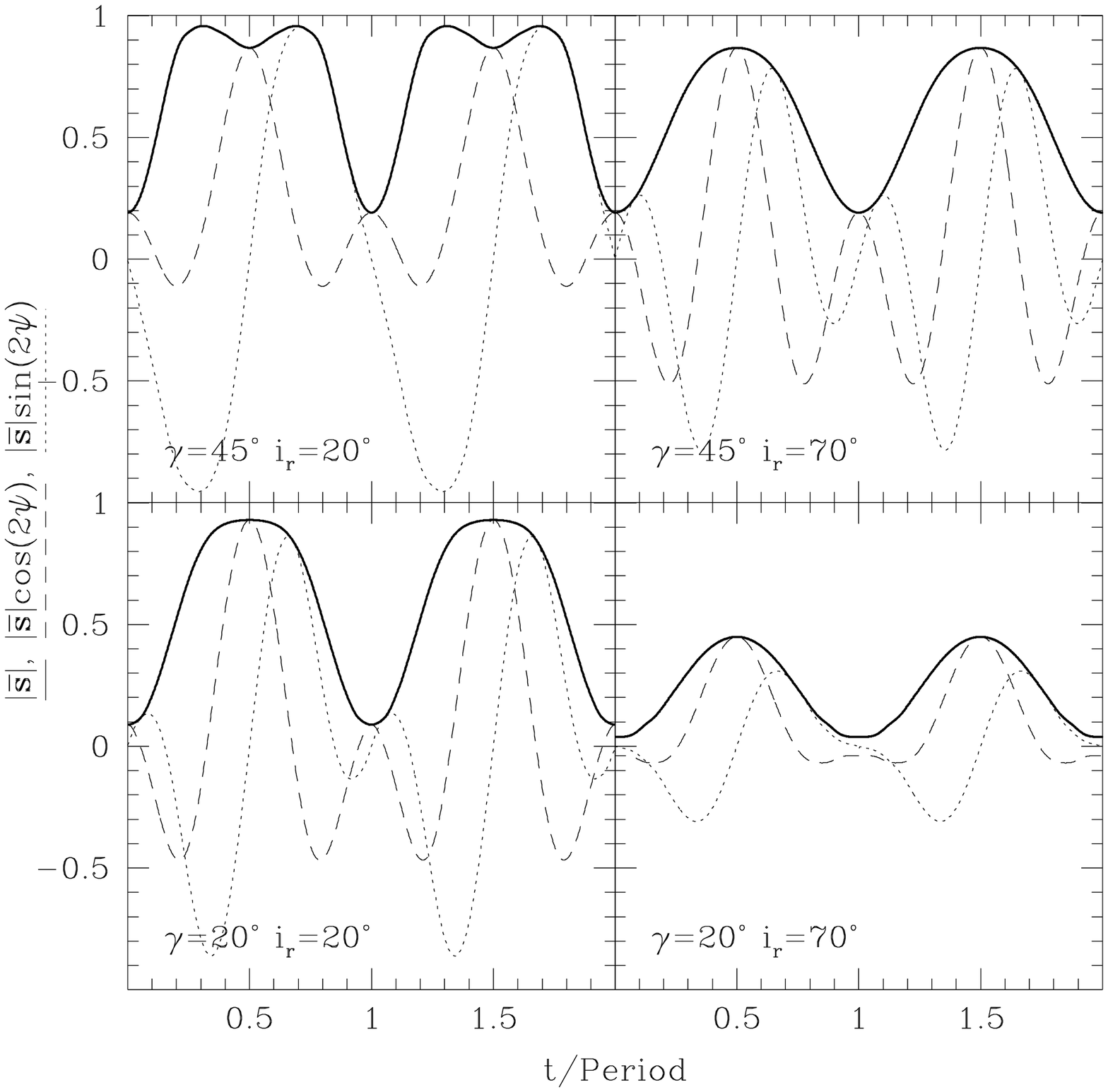}
 \caption{The polarized light curve for $\nu \mu_{30}^2=10^{15}$ Hz
 and $R=10$ km. }
\label{fig:lc15}
\end{figure}

\begin{figure}
\plotone{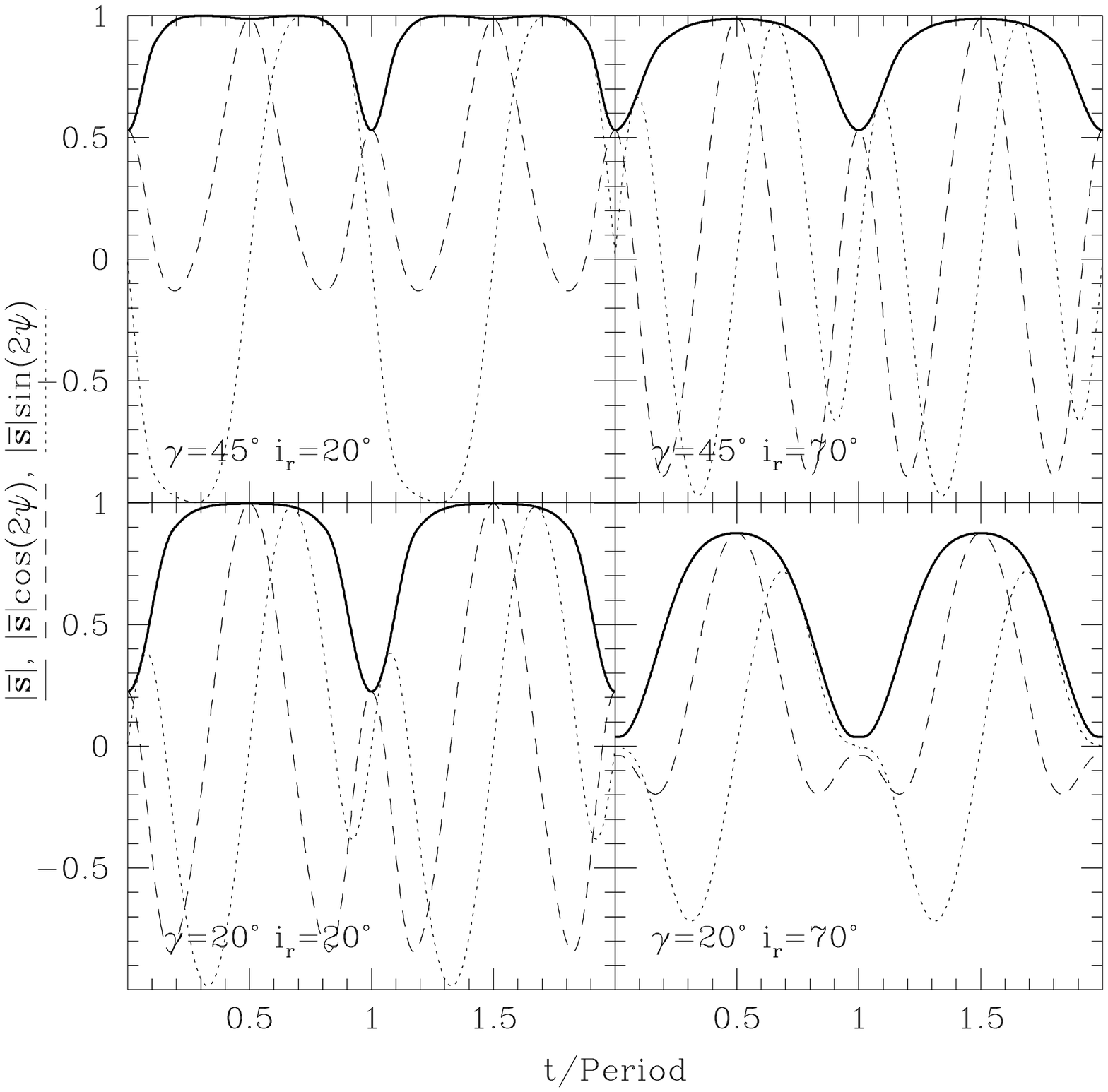}
 \caption{The polarized light curve for $\nu \mu_{30}^2=10^{17}$ Hz
 and $R=10$ km. }
\label{fig:lc17}
\end{figure}

 Results for the ``polarization light-curve'' for several frequencies
 and several NS angles are depicted in
 figures~\ref{fig:lc0}-\ref{fig:lc17}. The solid curve is the total
 polarization. It could be calculated if the measurement yields both
 $S_{O,1}$ and $S_{O,2}$. If however a polarimeter just measures one
 axis, and it is aligned with $y$ then its measurements will follow
 the dashed line. If rotated by $45^{\circ}$, the data will follow the
 dotted line. Clearly, the best possible measurement is that of the
 time behavior of the polarization in two axes which yields the large
 net polarization $|\bar{S}|$ and the various angles in the
 system. The latter include the angle separating the axes $\gamma$,
 the observer's inclination above the rotational equator $i_r$, as
 well as the direction of the rotational axis in the sky.

\subsubsection{Time averaged polarization of a NS}
\label{sec:timeaverage}

 Although the best polarization measurement possible should be time
 resolved, often is it hard to do so.  It is easier to measure the
 polarization averaged over the spin period.  By looking at
 figures~\ref{fig:lc0}-\ref{fig:lc17}, we see that if our polarimeter
 is aligned with the rotational axis, a net polarization signal is
 obtained though it is typically significantly smaller than the
 absolute polarization. If the polarimeter is rotated by $45^{\circ}$,
 symmetry dictates a null average signal.

 The average polarization depends on $\gamma$ -- the separation
 between the axis, and $i_r$, the inclination above the rotational
 equator. Although for the general population of NS, the two should
 not be correlated, this is not the case if we wish to study NSs for
 which their geometry is already known. NS geometry is known for some
 pulsars from linear polarization swing measurements in the
 radio. Because the objects have to be pulsars with beams passing close
 to the line of sight, there is a selection effect which chooses
 objects with only $i_r \sim \pi/2 - \gamma$.

\begin{figure}
\plotone{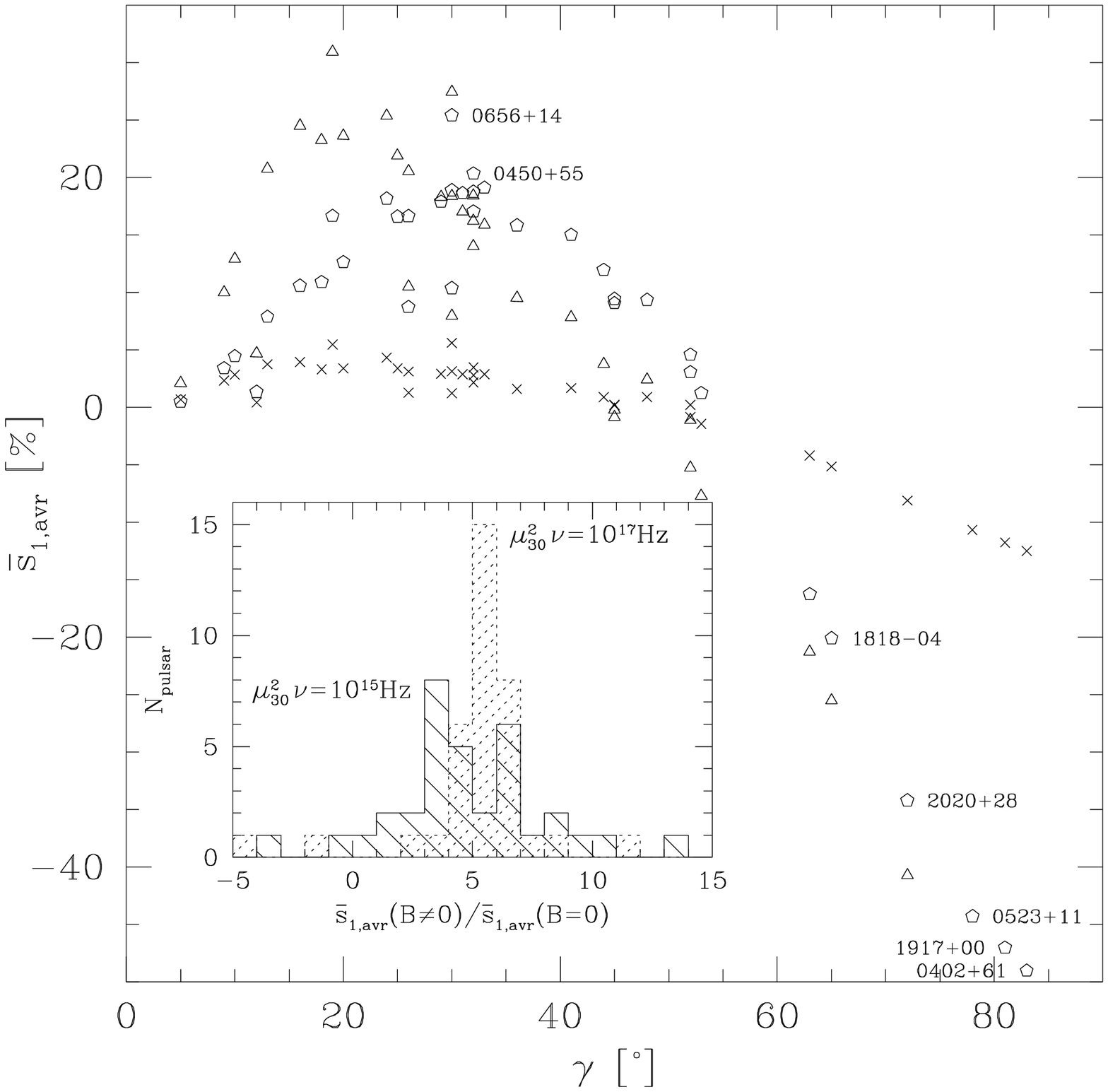}
 \caption{The average linear polarization expected for the thermal
 radiation of pulsars for which their geometrical angles are known
 ($\gamma$ - the separation between the magnetic and rotation axes,
 and $i_r$ the observer's inclination above the rotational
 equator). The crosses are the small polarizations expected if
 polarization dragging is neglected. The triangles and pentagons are
 the higher polarizations expected for $\nu \mu_{30}^2=10^{15}$ and
 $10^{17}$~Hz respectively when $R=10$~km (\ie , for optical and X-ray
 frequencies, for a typical $10^{12}$~G NS. The inset is a histogram of
 the increase in average polarization when QED is not neglected. }
\label{fig:avr_pol}
\end{figure}

 \figref{avr_pol} describes the average expected polarization for
 the different pulsars for which $\gamma$ and $i_r$ are known. The
 data are taken from \jcite{mitra}. We find that the expected time
 averaged polarization for the thermal radiation of $10^{12}$~G type
 pulsars is going to be on average 5-7 times larger if QED effects are
 properly taken into account and measurement is done in the optical or
 X-ray. One example is PSR~0656+14 which has a measured thermal
 spectrum (\cite{Pavl97}). The prediction is that the time average of
 its polarization is going to be about 25\% times the typical
 intrinsic polarization of an average surface element.  This should be
 compared with a 5\% prediction times the typical average intrinsic
 polarization, if polarization dragging and aligning does not take
 place. Note that even if the surface elements were to emit completely
 polarized radiation, then the maximum possible time average
 polarization that can be obtained for any geometrical configuration
 is 12.5\% if QED is neglected with typical values being significantly
 smaller. Thus, the time average measurement of the polarization of
 PSR~0656+14 is sufficient to prove the effects of QED on aligning the
 polarization. Because the pulsar was detected in optical and UV,
 polarimetry can in principle already be done with very long
 observations.

 We must be careful not to overstate the observability of this effect
 in the optical and near ultraviolet.  Typically, if thermal emission
 from the surface of the star dominates in the optical, the sources
 are exceptionally faint and would require approximately one night of
 observing time on a ten-meter-class telescope to detect the intrinsic
 polarization of the source.  Furthermore, contamination by
 non-thermal emission is typically important.  For example, even in
 the optical $\sim 30\%$ of the emission from PSR~0656+14 is
 non-thermal (\cite{Pavl97}).  Disentangling these two emission
 mechanisms in the optical is difficult but possible in principle.  In
 the X-rays the signal is much stronger and non-thermal emission plays
 a lesser role.  However, we do not now have instruments measure the
 polarization of X-ray radiation from astrophysical sources.

\section{Discussion \& Summary}
\label{sec:discussion}

 It is well known that the intrinsic polarization of the thermal
 radiation emanating from any NS surface element should be highly
 polarized. This is a direct result of the effects that the magnetic
 field has on photon propagation. However, it was thought until
 recently that because each surface element has a different magnetic
 field orientation, the combined emission for all the different
 surface elements would result with a low net polarization for the
 integrated light. This conclusion, however, rests on the assumption
 that nothing special happens to the polarization angles along the
 way. Here, we have shown that QED does have a major effect on the
 polarization angles in the magnetosphere.

 In the presence of QED, the vacuum becomes birefringent. If the
 differences in the indices of refraction are large enough, `adiabatic
 evolution' of the polarization will evolve each polarization state
 separately up to a distance $r_\rmscr{pl}$, the polarization-limiting
 radius. If the states change slowly because the magnetic field 
 orientation changes, the direction of polarization will change as
 well. This phenomenon is known to be important for radio waves due to
 plasma birefringence (\cite{Chen79}). The main differences between
 the two effects are first that plasma birefringence becomes
 progressively more important for long wavelengths, as opposed to the
 shorter wavelengths in which vacuum birefringence becomes
 progressively more important. Second, the amount of actual plasma
 birefringence is hard to predict accurately, because the amount of
 plasma present varies according to the pulsar model adopted. Vacuum
 birefringence depends only on the magnetic field of the NS.

 If the polarization limiting radius $r_\rmscr{pl}$ is far from the
 surface of the NS, the adiabatic evolution arising from QED
 birefringence has a very interesting effect---it aligns the
 polarization angles such that large net polarizations are
 obtained. The further from the surface that the coupling of the modes
 takes place (where adiabatic evolution fails), the better is the
 alignment of rays originating from different surface elements. For
 typical magnetic fields of $10^{12}$~G, the alignment is already
 important for optical and UV photons. And it should be almost
 complete in X-rays. In stronger fields, as are predicted to exist on
 magnetars, the alignment should be almost complete even in the
 optical and the polarization would be very high. This should be
 compared with the predictions neglecting polarization alignment which
 always result with significantly lower polarizations.

 If the magnetic and rotational axes are misaligned, as is generally
 the case, the direction of linear polarization changes with the
 rotation phase. As a result, the time-averaged measurements generally
 yield smaller net polarizations than time-resolved measurements. The
 latter are therefore much more preferable. However, they require a
 more elaborate measuring technique which for the very faint thermal
 signal of PSRs is highly nontrivial.

 Generally, a circular polarization component along one ray arises
 when the polarization limiting radius is not orders of magnitude
 larger than the radius of the NS. This is due to the fact that while
 coupling of the states occur at $r\sim r_\rmscr{pl}$ the magnetic
 field is changing its orientation relative to the ray.  However, when
 summed over the image, the circular component vanishes by symmetry.
 Therefore, any measurement of a non vanishing circular component in
 the thermal radiation would imply that the system has broken its
 symmetry between the apparent sides of the magnetic axis. This can
 happen for example if the magnetic field has a non symmetric
 component (rotation, higher multipoles, offset dipole, etc.).  It
 can happen if the temperature (and therefore emission) is not only a
 function of magnetic latitude (e.g., if there are `hot spots'). It
 can also arise because a rotating NS will Doppler boost the radiation
 from one apparent side of the rotation axis to the blue and the other
 side to the red. A circular component was also shown to arise when
 taking the effects that rotation have on the decoupling process
 itself (\cite{Heyl99polar}). The main difference between the two
 types of circular components is that the latter type increases with
 frequency, while the circular component that arises from asymmetries
 is largest for optical or UV (for $\sim 10^{12}$~G), when the
 polarization limiting radius is comparable to the radius of the NS.

 Polarization measurements of the thermal radiation will clearly be
 very beneficial. First, the measurement of polarization will verify
 the birefringence induced by a magnetic field predicted by QED.
 Magnetic vacuum birefringence has not yet been detected.  Moreover,
 measurement of polarization often elucidates the geometry of the
 systems. In this case however, it could also give information on
 actual physical parameters. For example, if the magnetic dipole
 moment $\mu$ is known (\eg from spin-down rate measurement) then $R$
 can be extracted from the polarization measurement which indicate how
 much alignment has taken place. The more alignment observed, the
 smaller the radius has to be because the NS apparent solid angle at
 $r_\rmscr{pl}$ is then smaller.  In AXPs, it could be used to verify
 their extreme magnetic nature.

\acknowledgments
Support for this work was provided by the National Aeronautics and
Space Administration through Chandra Postdoctoral Fellowship Award
Number PF0-10015 issued by the Chandra X-ray Observatory Center, which
is operated by the Smithsonian Astrophysical Observatory for and on
behalf of NASA under contract NAS8-39073. NS wishes to thank CITA for
the fellowship which supported him. 

\def\mn{MNRAS}

\bibliographystyle{jer}
\bibliography{mine,physics,ns,gr,polarpaps}

\end{document}